\documentclass{WileyMSP-template}
\usepackage{setspace}
\usepackage{microtype}
\microtypesetup{protrusion=true,expansion=true}
\usepackage{cite}
\usepackage{amsmath} 
\usepackage{bm} 
\usepackage{xcolor}

\begin{document}

\pagestyle{fancy}
\rhead{\includegraphics[width=2.5cm]{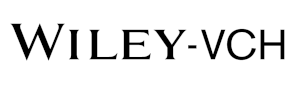}}

\title{Mesoscale Domain Evolution Mechanism during  {Alternating} Current (AC) Poling of Relaxor Ferroelectrics}

\maketitle


\author{Yuan-Jie Sun}
\author{Bo Wang\textsuperscript{*}}
\author{Long-Qing Chen\textsuperscript{**}}


\begin{affiliations}
Yuan-Jie Sun, Dr. Bo Wang and Prof. Long-Qing Chen\\
Department of Materials Science and Engineering and Materials Research Institute, The Pennsylvania State University, University Park, PA 16802, USA\\
\textsuperscript{*}bzw133@psu.edu; \textsuperscript{**}lqc3@psu.edu

\end{affiliations}



\begin{abstract}

{Ferroelectric domain variants that are energetically equivalent are expected to remain preserved during polarization reversal under a symmetry-preserving electric field. However, recent experiments on relaxor-ferroelectric crystals have revealed irreversible elimination of inclined domain walls during AC poling, while the underlying mesoscale mechanism remains unclear.} Here, we investigate domain-wall motion during AC poling of rhombohedral Pb(Mg$_{1/3}$Nb$_{2/3}$)O$_3$--PbTiO$_3$ single crystals containing 71$^\circ$ and 109$^\circ$ domain walls {within a quasi-two-dimensional laminated geometry} using phase-field simulations. Simulations reveal that domain-wall behavior during polarization reversal depends on the spacing ratio between 71$^\circ$ and 109$^\circ$ domain walls. Closely spaced 71$^\circ$ domain walls undergo irreversible elimination, whereas widely separated walls are preserved. A threshold ratio for domain-wall elimination is identified and found to depend on mechanical boundary conditions. {By tracking domain-wall trajectories, we attribute this behavior to unsynchronized motion of neighboring 71$^\circ$ domain walls arising from long-range elastic interactions when walls become strongly coupled. This collective motion breaks the symmetry between domain variants and leads to irreversible domain-wall elimination.} {These findings provide mechanistic insight into collective domain-wall evolution during polarization reversal and suggest that proximity-driven symmetry breaking may provide a mesoscale mechanism for domain engineering in ferroelectrics with high domain-wall densities.}
\end{abstract}


\section{Introduction}

Ferroelectric materials are characterized by the presence of polar domains with well-aligned electric dipoles that can be reoriented by applying an electric field. For many functional applications, ferroelectric materials must first be polarized to create a macroscopic net polarization. This poling process typically proceeds through the nucleation and growth of energetically favored domain variants at the expense of unfavorable ones, mediated by domain-wall motion. For ferroelectric single crystals poled along a crystallographic polar axis, a sufficiently strong electric field can stabilize a nearly single-domain state. In contrast, when the poling field is applied to a nonpolar axis, multidomain configurations emerge, consisting of symmetry-equivalent domain variants that share identical free energies under the applied field. By tuning the direction~\cite{park1997ultrahigh}, magnitude~\cite{kim2024electrical}, and waveform~\cite{sun2021dielectric} of the poling field, a wide variety of domain configurations can be engineered in a ferroelectric, enhancing functional properties such as dielectric permittivity~\cite{chang2018dielectric}, piezoelectricity~\cite{qiu2020transparent}, and electro-optic properties~\cite{liu2024large} that are essential for energy storage~\cite{cheng2017demonstration,tyagi2026pressure, singh2025v2se2o}, ultrasound transduction~\cite{zhang2012high}, and optical modulation applications~\cite{liu2022ferroelectric}.

Domain engineering via advanced poling protocols has been especially successful in the development of relaxor ferroelectric single crystals~\cite{qiu2020transparent,liu2022ferroelectric,finkel2022simultaneous,negi2023ferroelectric,xu2026cryogenic}. Relaxor ferroelectrics are characterized by hierarchical domain structures with domain sizes spanning multiple length scales that contribute to macroscopic physical properties synergistically~\cite{otonicar2018multiscale}. A prominent example is found in $[001]$-poled rhombohedral relaxor-PbTiO$_3$ crystals, such as Pb(Mg$_{1/3}$Nb$_{2/3}$)O$_3$--$x$PbTiO$_3$ (PMN--PT)~\cite{zhang2012high}. Comprehensive enhancement of dielectric, piezoelectric, and optical properties has been demonstrated in these materials systems using advanced poling techniques, most notably by alternating current (AC) poling~\cite{kim2022areview,maiwa2025microstructure,sun2022recent}. The improved physical properties have been attributed to the modification of phase symmetries and domain structures in these crystals by AC poling~\cite{kim2022areview}. Conventional poling technique based on a direct-current (DC) electric field along the $[001]$ direction stabilizes a four-variant rhombohedral (4R) domain structure with a lamellar morphology, whereas AC poling has been shown to further eliminate two of the rhombohedral variants, resulting in a two-variant rhombohedral (2R) domain state with layered domain structures~\cite{qiu2020transparent,qiu2021insitu}. However, all rhombohedral domain variants of the DC-poled 4R configuration, by symmetry, possess identical bulk free energies under an antiparallel electric field applied along $[00\bar{1}]$. As a result, antiparallel poling during AC poling does not break the symmetry among the domain variants and shall not generate a bulk thermodynamic driving force favoring one variant over another. But why are certain domains irreversibly eliminated during AC poling when all domain variants are energetically equivalent? 

This apparent discrepancy raises a fundamental question that remains unresolved, largely due to the experimental difficulty of capturing domain and domain-wall behavior during poling with sufficiently high temporal and spatial resolution. Considerable experimental efforts have been devoted to characterizing the complex hierarchical domain morphologies and electric-field-driven domain and domain-wall dynamics in relaxor-PT systems~\cite{liu2019realtime,sato2024response,zhang2023varied,perez-moyet2023anisotropy}. However, the real-space domain dynamics during poling are still not well understood~\cite{qiu2021insitu,sato2024response,sun2025enhanced,brichta2025quantifying,ushakov2021domain}. In parallel, computational modeling~\cite{yang2024first,yang2025terahertz}, particularly the phase-field method, has been employed to study domain evolution under applied electric fields~\cite{qiu2020transparent,wan2021alternating}. These simulations have shown that 71$^\circ$ domain walls inclined to the $[001]$ poling-field direction can be fully eliminated over multiple cycles of AC poling, with the elimination efficiency influenced by the field frequency and mechanical boundary conditions. Although irreversible domain-wall elimination during AC poling has been reproduced in simulations, the underlying mesoscale mechanisms remain elusive.

In this work, we address this fundamental question by systematically investigating domain-wall dynamics during polarization switching in the $[001]$-poled rhombohedral PMN--PT single crystal subjected to an antiparallel electric field using phase-field simulations. By employing quasi-2D laminated domain structures with controlled initial domain-wall spacing, we identify the critical role of relative domain size in determining whether adjacent 71$^\circ$ domain walls undergo irreversible annihilation or are reversibly preserved during polarization reversal. We further reveal that the domain-wall elimination results from asymmetric and unsynchronized domain-wall dynamics, which is governed by proximity-induced correlation effects via imbalanced elastic interactions.  {These findings provide mechanistic insight into domain evolution during the antiparallel-switching stage of AC poling in laminated rhombohedral PMN--PT and highlight collective domain-wall interactions as an important factor governing polarization reversal in multidomain ferroelectrics.}

\section{Results}

\begin{figure}[t]
\centering
\includegraphics[width=0.4\linewidth]{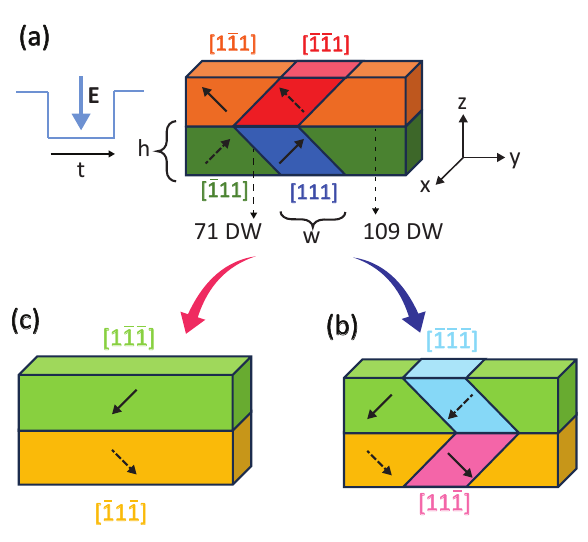}%
\caption{ {Schematic illustration of domain-configuration evolution under an antiparallel electric-field pulse.
(a) Initial $[001]$-poled four-variant rhombohedral (4R) domain configuration containing both 109$^{\circ}$ and 71$^{\circ}$ domain walls (DW). The domain height $h$ and width $w$ are defined for a domain bounded by a pair of adjacent 71$^{\circ}$ domain walls.
(b) Preservation of the 71$^{\circ}$ domain wall after polarization reversal, giving a 4R configuration topologically equivalent to the initial state.
(c) Elimination of the 71$^{\circ}$ domain wall after polarization reversal, giving a two-variant rhombohedral (2R) configuration.}}
\label{fig:1}
\end{figure}

It has been well established that $[001]$-poled rhombohedral PMN--PT single crystals by conventional DC electric field exhibit a laminated domain structure composed of four symmetry-equivalent rhombohedral domain variants with polarization vectors pointing to $[\pm1, \pm1, 1]$ separated by 71$^\circ$ and 109$^\circ$ domain walls~\cite{qiu2021insitu,xiong2022performance}. As schematically illustrated in Figure~\ref{fig:1}(a), the 109$^\circ$ domain walls are oriented parallel to the $(001)$ planes, forming a layered structure consisting of alternating in-plane polarization. Within each layer, a set of 71$^\circ$ domain walls associated with $\{101\}$ planes are present, inclined with respect to 109$^\circ$ domain walls by an angle of approximately 45$^\circ$. Owing to the crystallographic symmetry of the rhombohedral phase, all four polarization variants in the 4R structure are energetically equivalent under DC poling along $[001]$ or $[00\bar{1}]$.

 {During AC poling, the antiparallel half-cycle is the critical stage at which 71$^\circ$ domain-wall motion and possible elimination are expected to occur~\cite{qiu2020transparent}.} When an antiparallel electric field is applied along the $[00\bar{1}]$ direction with a  {sufficiently} high magnitude, the polarization states of the pre-poled 4R states become unstable and tend to be reversed.  During this process, two distinct behaviors of the 71$^\circ$ domain walls have been reported. In one case, the 71$^\circ$ domain walls are preserved after polarization reversal, but reorient by swinging between crystallographically equivalent planes, such as from $(101)$ to $(10\bar{1})$, while maintaining the overall laminated 4R topology, as illustrated in Figure~\ref{fig:1}(b). In the other case, the 71$^\circ$ domain walls are irreversibly eliminated, resulting in a reduced two-rhombohedral-variant (2R) domain structure composed exclusively of 109$^\circ$ domain walls, as shown in Figure~\ref{fig:1}(c). The coexistence of these two outcomes under nominally similar antiparallel poling conditions highlights a fundamental unresolved question: why symmetry-equivalent 71$^\circ$ domain walls are sometimes preserved and sometimes eliminated during polarization reversal.

We begin by demonstrating the two scenarios illustrated in Figure~\ref{fig:1} of domain-wall elimination or preservation using phase-field simulations. Inspired by previous studies~\cite{qiu2020transparent}, we hypothesize that the initial domain width-height ratio, $w/h$, of the laminated 4R domain structure plays a critical role in determining the domain evolution during polarization reversal. The domain height $h$ and width $w$ are defined in Figure~\ref{fig:1}(a). We constructed a series of 4R domain structures spanning a range of $w/h$ ratios and relaxed each configuration to equilibrium using phase-field simulations to represent the initial DC-poled states. An antiparallel electric-field pulse was then applied for a sufficiently long duration, and the domain evolution was tracked until no further apparent changes were observed. The electric field was subsequently removed, and the system was allowed to relax to a new equilibrium.  {We note that this single antiparallel pulse represents the switching half-cycle of an AC poling sequence. We adopted this simplified protocol to isolate and focus on the elementary mesoscale domain-wall evolution mechanism dynamics during polarization reversal, rather than a systematic study on the effects of waveform, frequency, and repeated cycling.} The detailed modeling formulation and simulation parameters are provided in the Methods section. 

\begin{figure}[t]
\centering
\includegraphics[width=0.8\linewidth]{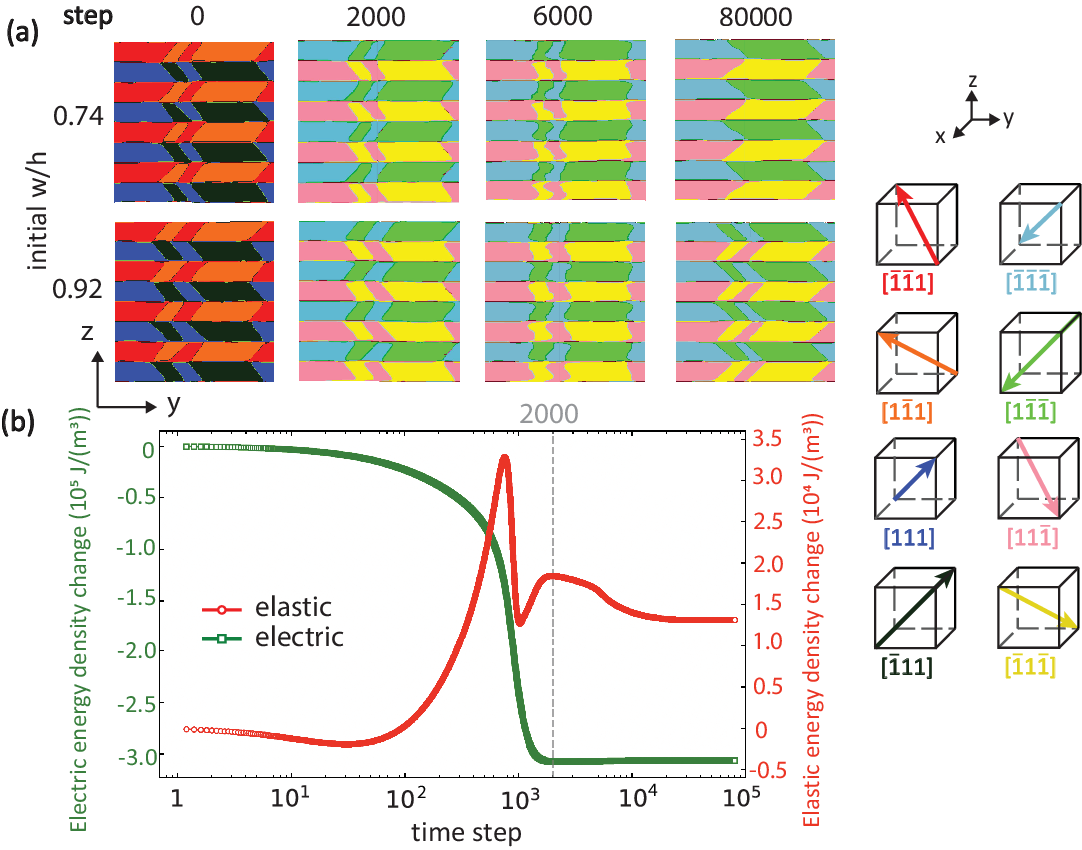}%
\caption{Phase-field simulation of poling dynamics and associated energy evolution in PMN--PT single crystals.
(a) Domain evolution under a mechanical clamping condition, $\langle\bm{\varepsilon}\rangle=0$, shown for two aspect ratios, $w/h=0.74$ and $0.92$, at simulation steps 0, 2000, 6000, and 80000.
(b) Temporal evolution of the electric and elastic energy contributions during poling for $w/h=0.74$ case in (a). The initial stage (0–2000 steps) is dominated by a rapid decrease in electric energy, corresponding to polarization switching. The subsequent stage (2000 steps to the end) exhibits a gradual reduction in elastic energy, corresponding to the domain-wall relaxation and motion.}
\label{fig:2}
\end{figure}

Figure~\ref{fig:2} summarizes the evolution of domain morphology during and after application of the antiparallel electric-field pulse for two representative cases, $w/h = 0.74$ and $w/h = 0.92$. In both scenarios, two characteristic stages of domain morphology evolution are observed. In the first stage, polarization reversal begins with the collective rotation of the local polarization vectors in each domain variant (e.g., from $[111]$ to $[11\bar{1}]$) without breaking the topology of the laminated 4R structure, leading to reversal of the net polarization along the out-of-plane direction. The subsequent evolution of the 71$^\circ$ domain walls differs markedly in the second stage. For the smaller aspect ratio ($w/h = 0.74$), the pair of adjacent 71$^\circ$ domain walls is irreversibly eliminated at this stage of switching. In contrast, for the larger aspect ratio ($w/h = 0.92$), the 71$^\circ$ domain walls are largely preserved after polarization reversal but switch to an energetically equivalent orientation. Note that the 109$^\circ$ domain walls lying horizontally within the (001) plane remain intact during the whole poling process.

The two characteristic stages of domain morphology evolution can be rationalized by examining the electrostatic and elastic energy contributions during switching (Figure~\ref{fig:2}(b)). In the first stage (0–2000 timesteps), the electrostatic energy decreases rapidly at the expense of increasing elastic energy, consistent with local polarization reversal under the applied field and the transient formation of domain walls along mechanically incompatible orientations that generate large elastic strains. In the second stage (2000 timesteps and beyond), the elastic energy relaxes as the domain walls reorient toward mechanically compatible configurations while the electrical energy remains unchanged. For instance, the 71$^\circ$ domain wall between domain variants with $\mathbf{P} \parallel [11\bar{1}]$ and $\mathbf{P} \parallel [1\bar{1}\bar{1}]$ rotates from the $(101)$ to the $(10\bar{1})$ plane. 

Notably, Figure~\ref{fig:2}(b) shows the energy evolution only for the case exhibiting domain-wall elimination ($w/h = 0.74$). The corresponding plot for domain-wall preservation ($w/h = 0.92$) is provided in Figure S1. The two cases display nearly identical energy variations during switching, indicating that the macroscopic energy evolution alone does not qualitatively distinguish domain-wall elimination from preservation.  {Nevertheless, after the applied field is removed and the system relaxes to equilibrium, the total energy decreases when domain-wall elimination occurs, whereas it increases slightly when the domain walls are preserved.} The energy reduction in the former is primarily attributable to the decrease in gradient energy associated with the reduced domain-wall density. The slight energy increase in the latter is related to the quantitative change in $w/h$ following polarization reversal, as discussed later.

\begin{figure}[t]
\centering
\includegraphics[width=0.75\textwidth]{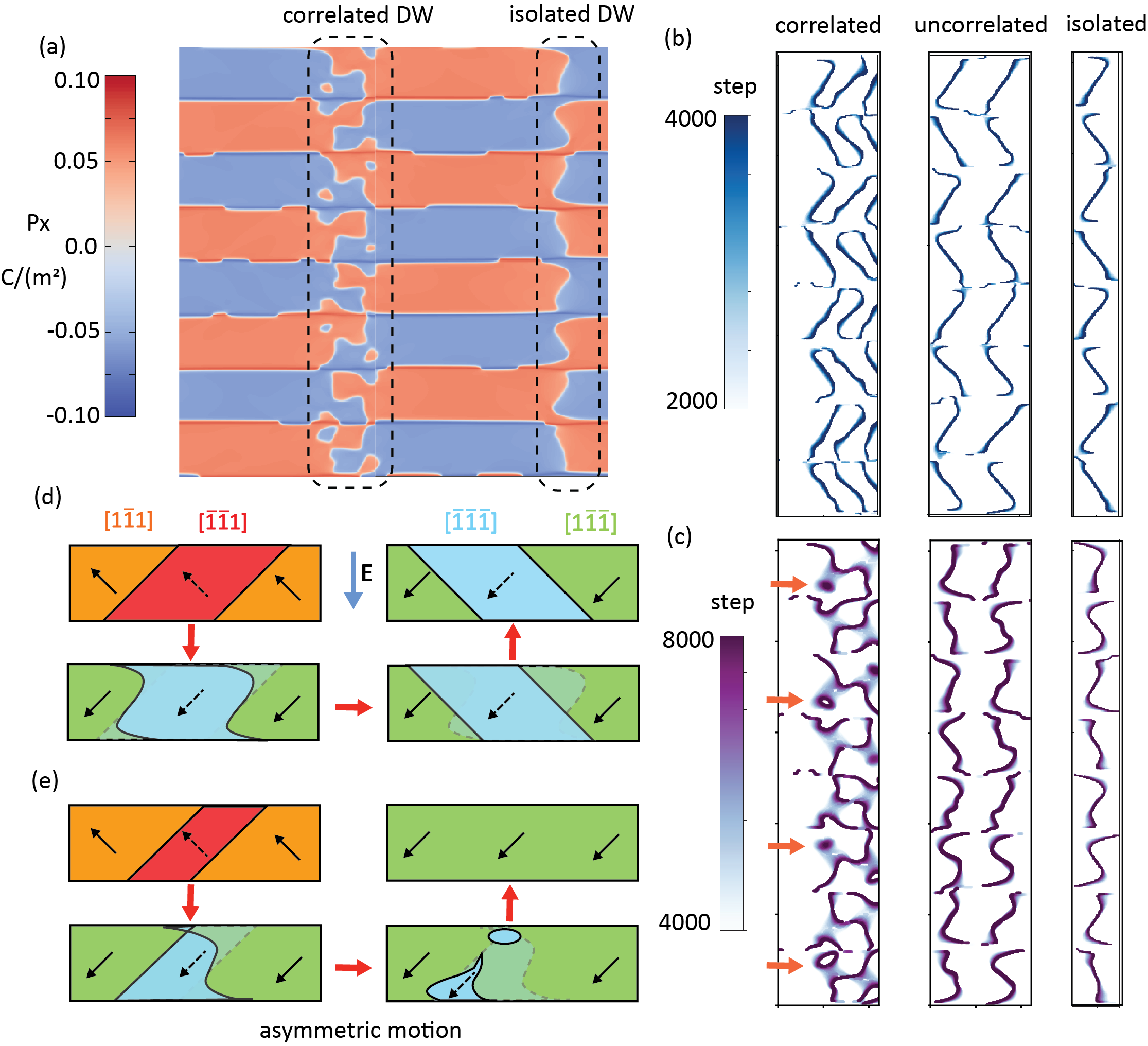}%
\caption{Distinct modes of domain-wall motion during antiparallel poling in PMN--PT. (a) Snapshot of the polarization component $P_x$ at step 8000 for $w/h=0.74$ under fixed-strain boundary condition ($\langle\epsilon\rangle=0$), illustrating two characteristic behaviors: correlated domain walls (DWs) and isolated DWs.
(b,c) Trajectories for correlated ($w/h=0.74$), uncorrelated ($w/h=0.92$), and isolated ($w/h=3.26$) DWs from (b) steps 2000--4000 and (c) steps 4000--8000. The orange arrows in (c) highlight formation of bubble-like nanodomains during the evolution of correlated DWs. (d) Schematic illustration of symmetric and synchronized behavior of uncorrelated DWs during switching.
(e) Schematic illustration of asymmetric and unsynchronized behavior of  {correlated} DWs during switching.
}
\label{fig:3}
\end{figure}

To uncover the mesoscale origin of the two distinct modes of domain-wall behavior, the trajectories of individual domain walls during the polarization switching are examined. The domain-wall trajectories are extracted using the $P_x$ component of polarization as a marker (Figure~\ref{fig:3}(a)), since $P_x$ is the only polarization component that reverses across a 71$^\circ$ domain wall in the 4R configuration. Based on the spatial separation  {and interaction strength} of the domain walls, three representative cases can be identified: (i) an isolated 71$^\circ$ domain wall that is well separated from other walls; (ii) a pair of uncorrelated domain walls that are spatially adjacent but weakly interacting (as in the $w/h$ = 0.92 case), and (iii) a pair of correlated domain walls that are closely spaced and strongly interacting (as in the $w/h$ = 0.74 case).  {The motion of an isolated wall is governed mainly by its local elastic and electrostatic environment rather than by direct coupling to another nearby inclined wall. Uncorrelated wall pairs exhibit approximately symmetric motion, so the corresponding domain may shrink but does not disappear within one switching step. In contrast, correlated wall pairs undergo unsynchronized motion which leads to elimination within one antiparallel switching step.} Our analysis focuses on the time window from 2000 to 8000 timesteps, corresponding to the early part of the elastic relaxation stage during which the energy varies most significantly (cf. Figure 2b).

Figure~\ref{fig:3}(b) shows the domain-wall trajectories during the early part of elastic relaxation (2000--4000 timesteps). We found that isolated domain walls and uncorrelated domain-wall pairs exhibit synchronized motion, with the endpoints of each 71$^\circ$ domain-wall segment deviating slightly from their initial positions near junctions with the bounding 109$^\circ$ domain walls. This symmetric response gives rise to smooth, S-shaped domain-wall trajectories. In contrast, correlated domain-wall pairs display unsynchronized motion, with one wall advancing more rapidly than the other and approaching neighboring 109$^\circ$ domain-wall junctions.

As elastic relaxation proceeds (4000--8000 timesteps), the qualitative difference between these behaviors becomes more pronounced, as shown in Figure~\ref{fig:3}(c). For correlated domain-wall pairs, the asymmetric domain-wall motion leads to the two adjacent 71$^\circ$ domain walls intersecting near the 109$^\circ$ domain-wall junctions, resulting in the formation of isolated bubble-like nanodomains detached from the surrounding laminated structure, as indicated by arrows in Figure 3(c). These nanodomains subsequently shrink and vanish, leaving no nucleus for the reversed domain variant and leading to irreversible elimination of the 71$^\circ$ domain. In contrast, for isolated and uncorrelated domain walls, the continued synchronized motion prevents intersection, allowing the 71$^\circ$ domains to survive and adjust toward new equilibrium orientations. These distinct behaviors are schematically summarized in Figure 3(d,e).

These observations demonstrate that irreversible domain-wall elimination arises from correlated domain-wall dynamics that break the symmetry between otherwise equivalent domain walls. When two adjacent 71$^\circ$ domain walls are sufficiently close to interact, their unsynchronized motion promotes domain-wall intersection and eventually irreversible domain-wall elimination, whereas isolated or weakly interacting domain walls evolve in a more synchronized manner and are therefore preserved after the switching step.

To quantify the critical condition under which domain-wall elimination occurs, we performed a series of phase-field simulations with systematically varied initial $w/h$ and measured its final value in the equilibrated state after the antiparallel poling. The results are summarized in Figure 4(a). From this parametric analysis, a critical initial aspect ratio of $w/h \approx$ 0.75 is identified, below which adjacent 71$^\circ$ domain walls are irreversibly eliminated. Near this threshold, an intermediate regime is observed in which the $w/h$ ratio is reduced after the antiparallel field poling but the 71$^\circ$ domain walls are not fully eliminated. The size-reduction behavior suggests progressive reduction of $w/h$ until reaching the threshold ratio and eventual elimination under repeated antiparallel switching. This possibility is further demonstrated for the case with initial $w/h$ = 0.94 (see Figure S2). These results provide a rationale for why it often takes a few cycles of AC fields to fully eliminate the 71$^\circ$ domain walls and achieve optimal piezoelectric responses in AC-poled relaxor ferroelectrics.

\begin{figure}[t]
\centering
\includegraphics[width=0.45\linewidth]{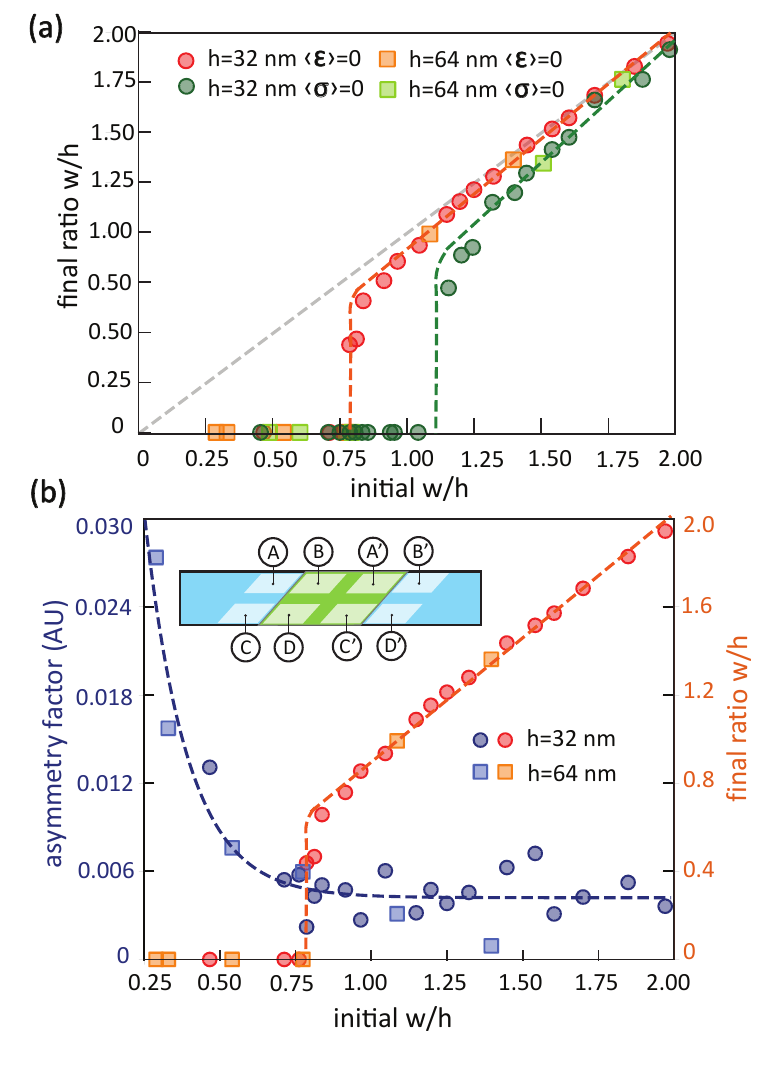}%
\caption{Evolution of $w/h$ ratio and domain-wall motion asymmetry. (a) Final $w/h$ as a function of the initial $w/h$ for  {lamellar heights} $h=32$ and $64$ nm under mechanical clamping $\langle\epsilon\rangle=0$ and stress-free condition $\langle\sigma\rangle=0$. $w/h=0$ indicates complete elimination of the 71$^\circ$ domain walls.  
(b) Dependence of the asymmetry factor (left axis) on $w/h$, plotted together with the resulting $w/h$. The inset schematically defines the sampling regions (A--D and A$'$--D$'$) used to evaluate the asymmetry factor.}
\label{fig:4}
\end{figure}

 {The analysis above is based on a single dimensionless parameter, $w/h$, which quantifies the relative size of the 71$^\circ$ domains. To test whether the criterion is scale free within the simulated range, we performed additional simulations in which the absolute lamellar height $h$ was varied. As shown in Figure~\ref{fig:4}(a), simulations with different lamellar heights ($h = 32$ and 64 nm) yield consistent threshold values of $w/h$, indicating that the elimination criterion is governed primarily by the geometric ratio $w/h$ rather than the absolute domain size.}

In addition, we examined the influence of mechanical boundary conditions by comparing simulations under zero-average-strain and zero-average-stress (Figure~\ref{fig:4}(a)). In both cases, proximity-induced domain-wall elimination is observed, indicating that the underlying mechanism does not depend on the specific mechanical constraint. Quantitatively, however, the critical threshold shifts, with stress-free conditions giving a larger threshold value ($w/h \approx 1$). This shift suggests a longer effective correlation length between adjacent 71$^\circ$ domain walls under stress-free conditions, consistent with reduced elastic constraints on domain-wall motion. This interpretation also provides a plausible explanation for why 2R layered structures are preferentially formed after AC poling cycles, as suggested in previous phase-field simulations~\cite{qiu2020transparent}. 

The energy analysis in Figure~\ref{fig:2}(b) suggests that the elastic energy contribution plays a dominant role in determining the asymmetric domain-wall behavior. Therefore, we introduce an asymmetry factor $\mathcal{A}$ to quantify the imbalance in elastic driving forces acting on adjacent domain walls, as shown in Figure~\ref{fig:4}(b). The definition of this factor is provided in Supporting Information Section IV.  {The spatial distribution of the shear-stress component dominating the elastic driving-force term is shown in Figure~S5 to better illustrate the asymmetry between neighboring correlated 71$^\circ$ domain walls.} Briefly, $\mathcal{A}$ measures the difference in elastic driving forces acting on a 71$^\circ$ domain by sampling local driving forces near the junctions of 71$^\circ$ and 109$^\circ$ domain walls on the upper (A and B) and lower (C and D) sides (see inset of Figure~\ref{fig:4}(b)). The asymmetry factor is evaluated by averaging multiple pairs of adjacent 71$^\circ$ domain walls at a snapshot prior to the onset of pronounced domain-wall motion (time 2000). The magnitude of $\mathcal{A}$ increases exponentially as $w/h$ decreases, correlating with the sudden drop of final $w/h$ (Figure~\ref{fig:4}(b)).  {This correlation supports the interpretation that unbalanced local elastic driving forces contribute to symmetry breaking between energetically equivalent domain variants during polarization reversal.}

\section{Discussion}

The present results reveal a mesoscale mechanism by which symmetry-equivalent domain walls can be irreversibly eliminated during polarization reversal, even in the absence of a macroscopic symmetry-breaking thermodynamic driving force. By combining time-resolved domain-wall tracing with quantitative domain-size analysis, we show that domain-wall elimination originates from a proximity-induced symmetry breaking between correlated neighboring domain walls.  {This mechanism provides a qualitative explanation for why nominally equivalent initial configurations can evolve into qualitatively different final states during AC poling. The idealized quasi-two-dimensional laminated 4R structures used in the simulations provide a controlled geometry for extracting the mesoscale mechanism, but they do not fully capture microstructural features of real relaxor ferroelectrics, such as defects, free surfaces, local chemical disorder, and multiscale three-dimensional domain morphologies~\cite{krogstad2018relation,eremenko2019local,zheng2024heterogeneous}. Further simulations incorporating these realistic features and larger simulation sizes will be needed to assess the quantitative importance of the proposed mechanism in experimental crystals.}

 {Recent experimental studies have provided direct evidence that AC poling can modify the domain configuration of relaxor-PT crystals through the elimination and reorganization of inclined domain walls ~\cite{bokov2026domain,perezmoyet2026insitu}. In particular, Bokov \textit{et al.} \cite{bokov2026domain} reported that AC poling of PMN--32PT crystals leads to the elimination of specific monoclinic ($M_B$) domain variants and produces intermediate poled states. Using rhombohedral PMN--38PT, Perez-Moyet \textit{et al.} \cite{perezmoyet2026insitu} further observed \textit{in situ} domain-wall merging dynamics during AC poling using polarized light microscopy, accompanied by an increase in the volume fraction of surviving rhombohedral domain variants. These experimental observations are broadly consistent with our conclusion that AC-related switching can reorganize the mesoscale domain-wall network through collective domain-wall evolution.} 

 {While it remains challenging to precisely control the domain aspect ratio $w/h$ of individual domains in experiments, previous studies have demonstrated the possibility of tuning the density of different types of domain walls through electrical, mechanical, and chemical approaches. For example, varying the pre-poling electric field and temperature, or using patterned electrodes, can modify the relative widths and densities of 71$^\circ$ and 109$^\circ$ domains~\cite{luo2020multilayered,negi2023ferroelectric}. Applying a pre-stress or mechanical confinement during the poling process can favor certain ferroelastic variants and thereby alter subsequent domain-wall motion~\cite{liu2019realtime,patel2015mechanical}. Chemical doping can also influence domain-wall configurations by modifying the intrinsic free-energy landscape and the energy barriers associated with domain-wall motion, as well as through changes in defect chemistry~\cite{bencan2020domain,schultheiss2023ferroelectric}. These effects may further influence the equilibrium density and stability of domain walls.}

Although the present study focuses on $[001]$-poled rhombohedral PMN--PT, the underlying mechanism is  {potentially relevant to other ferroelectric domain structures containing closely spaced inclined domain walls that separate symmetry-degenerate variants under an applied electric field. Similar domain configurations include $60^\circ$ domain walls in $[001]$-poled orthorhombic ferroelectrics and $90^\circ$ domain walls in $[111]$-poled tetragonal ferroelectrics, among others ~\cite{sun2014relaxorbased}. The proximity effect identified here may therefore provide a plausible mechanism for understanding domain evolution in domain-engineered ferroelectric crystals with high domain-wall densities.}

In addition, asymmetric behavior of a pair of adjacent domain walls has also been observed in ferroelectric thin films~\cite{jablonski2016asymmetric,zhang2019intrinsic}. For example, 71$^\circ$ domain walls in rhombohedral BiFeO$_3$ thin films have been reported to exhibit intrinsically asymmetric responses under opposite in-plane field directions~\cite{jablonski2016asymmetric}. Moreover, the domain-wall interactions have been reported to account for alternating conductive behavior between neighboring 71$^\circ$ domain walls in BiFeO$_3$ thin films~\cite{zhang2019intrinsic}.  {These observations suggest that the proximity-induced symmetry breaking may also be relevant to a broader class of ferroelectric materials.}

The present simulations employ antiparallel electric-field pulses to probe the dynamics of domain walls during polarization reversal, which is directly relevant to recent experimental demonstrations of pulse poling~\cite{xiong2022performance,mervosh2026far}. Such protocols have produced extraordinary piezoelectric performance and unusual aging behavior in doped relaxor-PbTiO$_3$, which has been attributed to a far-from-equilibrium state, yet its mesoscale origin remains unclear~\cite{mervosh2025far}. The transient intermediate states characterized by S-shaped domain walls (cf. Figure~\ref{fig:3}) might suggest a possible pulse-driven far-from-equilibrium state. While these states are transient in the present study, they are expected to be stabilized in doped relaxor-PT systems, where charged defects can serve as pinning centers of domain walls. 

In addition, experimental AC poling protocols have employed tailored waveforms with controlled amplitudes, frequencies, and the number of cycles, together with temperature control, to improve the functional performance of ferroelectric crystals~\cite{sun2022recent,kim2022areview}.  {We note that the present simulations adopt a single antiparallel electric-field pulse.}  However, our simulation framework and analysis can be extended to systematically investigate these poling parameters and establish quantitative process–domain structure relationships, which can be leveraged to deliberately tune domain-wall densities to optimize the dielectric, optical, and piezoelectric performance.  {We further note that the magnitude of the applied electric field can also alter the domain-evolution pathway. In particular, a sufficiently large electric field may drive a rhombohedral-to-tetragonal-like transformation and thereby remove tilted domain walls through a mechanism distinct from the proximity-induced mechanism discussed here. This field-induced phase-transformation pathway represents another possible mechanism beyond the scope of the present study but serves as an interesting topic for future work.} Future work incorporating more flexible and experimentally realistic poling protocols will be essential for translating the mechanistic insights obtained here into predictive guidelines for domain engineering in high performance ferroelectric materials.

\section{Methods}

Domain evolution in the ferroelectric crystal was simulated using the phase-field method based on Landau-Ginzburg-Devonshire (LGD) theory, as implemented in the $\mu$-PRO package. The polarization evolution is determined by solving the time-dependent Ginzburg–Landau (TDGL) equation~\cite{chen2008phase},  
\begin{equation}
   \frac{\partial P_i(\boldsymbol{x},t)}{\partial t}=-L\frac{\delta F}{\delta P_i(\boldsymbol{x},t)},
\end{equation}
where $P_i$ is ferroelectric polarization, $\boldsymbol{x}$ is position vector, $t$ is time, $L$ is the kinetic coefficient related to the domain-wall mobility, and $F$ is the free energy functional. It can be expressed as~\cite{chen2008phase,wang2019understanding}:

\begin{equation}
      F=\int_V[f_{\text{bulk}}(P_i)+f_{\text{grad}}(\partial P_i/\partial x_j)+f_{\text{elast}}(P_i, \varepsilon_{ij})
    +f_{\text{elec}}(P_i, E_i)] \text{d}x^3,
\end{equation}
    
where $f_{\text{bulk}}$, $f_{\text{grad}}$, $f_{\text{elast}}$, and $f_{\text{elec}}$ are bulk, gradient, elastic, and electric energy density, respectively.

The bulk energy density can be written as:
\begin{align}
f_{\mathrm{bulk}}(P_1,P_2,P_3,T)
&= \alpha_{1}\left(P_1^2+P_2^2+P_3^2\right)+ \alpha_{11}\left(P_1^4+P_2^4+P_3^4\right)  \nonumber\\
&\quad + \alpha_{12}\left(P_1^2P_2^2+P_2^2P_3^2+P_3^2P_1^2\right)
+ \alpha_{111}\left(P_1^6+P_2^6+P_3^6\right) \nonumber\\
&\quad + \alpha_{112}\Big[P_1^4\left(P_2^2+P_3^2\right)
+P_2^4\left(P_1^2+P_3^2\right)
 +P_3^4\left(P_1^2+P_2^2\right)\Big] + \alpha_{123}\,P_1^2P_2^2P_3^2,
\end{align}
where $\alpha_i$ denote the Landau coefficients relating to different-order dielectric stiffnesses.

The gradient energy density is determined by the polarization spatial gradient, i.e.,
\begin{equation}
f_{\mathrm{grad}}=\frac{1}{2}\,g_{ijkl}\,P_{i,j}\,P_{k,l},
\end{equation}
with $g_{ijkl}$ representing the gradient energy coefficients and $P_{i,j}=\partial P_i/ \partial x_j$.

The elastic energy density is defined by
\begin{align}
f_{\mathrm{elast}}
&= \frac{1}{2}\,c_{ijkl}\,e_{ij}(\mathbf{x})\,e_{kl}(\mathbf{x})
= \frac{1}{2}\,c_{ijkl}
\Big[\varepsilon_{ij}(\mathbf{x})-\varepsilon^{0}_{ij}(\mathbf{x})\Big]
\Big[\varepsilon_{kl}(\mathbf{x})-\varepsilon^{0}_{kl}(\mathbf{x})\Big],
\end{align}
with $c_{ijkl}$, $e_{ij}(\mathbf{x})$, $\varepsilon_{ij}(\mathbf{x})$, and $\varepsilon^{0}_{ij}(\mathbf{x})$
the elastic stiffness tensor, elastic strain, total strain, and eigenstrain tensors, respectively. The local eigenstrain $\varepsilon^{0}_{ij}(\mathbf{x})$ is related to the local polarization through
\begin{equation}
\varepsilon^{0}_{ij}(\mathbf{x}) = Q_{ijkl}\,P_k(\mathbf{x})\,P_l(\mathbf{x}),
\end{equation}
with $Q_{ijkl}$ the electrostrictive coefficients.

The electrostatic energy density is
\begin{equation}
f_{\mathrm{elec}}
= -P_i(\mathbf{x})\,E_i(\mathbf{x})
-\frac{1}{2}\,\varepsilon_0\,\kappa^{\,b}_{ij}\,E_i(\mathbf{x})\,E_j(\mathbf{x}),
\end{equation}
where $\kappa^{\,b}_{ij}$ is the background relative dielectric constant tensor and
$\varepsilon_0$ is the vacuum permittivity.

All material coefficients of rhombohedral PMN--PT used in the above equations were adopted from previous work~\cite{khakpash2015misfit}. The gradient energy coefficient is assumed isotropic and chosen to be $g_{11} = -g_{12} = g_{44} = 2.5199\times10^{-12}$ C$^{-2} $m$^4$ N, corresponding to a 180$^\circ$ domain wall with width 1 nm. 


The initial domain configuration of the simulations were preset to represent $[001]$-poled rhombohedral ferroelectric crystals with quasi-2D 4R domain structures. Four periods of 4R layered structures are assumed in the configuration to consider the statistical variations. This geometry was motivated by experimental observation~\cite{wan2021alternating,sun2025enhanced,qiu2020transparent} and allows for the convenience to define the layer height $h$ and domain width $w$. Periodic boundary conditions were imposed along $x$, $y$, and $z$. Two cell sizes were used, $1\times256\times256$~nm$^{3}$ and $1\times512\times512$~nm$^{3}$. Each sample comprised eight uniformly spaced layers separated by $109^{\circ}$ domain walls, with $h=32$~nm ($1\times256\times256$~nm$^{3}$ sample) or $64$~nm ($1\times512\times512$~nm$^{3}$ sample), to suppress stochastic single-layer effects. The domain aspect ratio was varied systematically over $w/h\in[0.26,2]$.

The pre-poled structures were first relaxed for $1.6\times10^{5}$ time steps. Subsequently, a uniform electric field was applied along $[00\bar{1}]$ with a constant magnitude exceeding the coercive field for $8.0\times10^{4}$ time steps.  {The antiparallel switching process spans approximately 0.74 ns when assuming $L\approx 4\times10^4$ C$^2$J$^{-1}$m$^{-1}$s$^{-1}$~\cite{hlinka2007mobility}. Because $L$ is a condition-dependent kinetic coefficient and has not been directly determined for the present system, this value should be regarded only as an approximate timescale estimate, while the qualitative results remain unchanged.} 
The electric field amplitude is set as $1.0\times10^{6}$~V/m under the zero-average-strain condition $\langle\bm{\varepsilon}\rangle=0$ and $5.0\times10^{5}$~V/m under the stress-free condition $\langle\bm{\sigma}\rangle=0$.  {Based on analytical estimates of the intrinsic coercive fields ($E_c$) under the corresponding mechanical boundary conditions, these field amplitudes are approximately $2.5E_c^{\langle\bm{\varepsilon}\rangle=0}$ and $2.9E_c^{\langle\bm{\sigma}\rangle=0}$, respectively. Therefore, both simulations were conducted within a comparable normalized field range to ensure complete polarization reversal.}  {Here, $\langle \bm{\varepsilon} \rangle =0$ represents a macroscopically clamped condition in which the sample is not allowed to change its average shape or size relative to the initial reference state, corresponding to the cubic paraelectric phase at the Curie temperature. Under this condition, the stress generated during polarization switching cannot be relaxed through macroscopic deformation of the sample, leading to the accumulation of elastic strain energy. In contrast, $\langle \bm{\sigma} \rangle =0$ denotes a macroscopically stress-free condition in which the average macroscopic stress of the crystal is allowed to relax to zero, although the local stress associated with polarization heterogeneities may remain finite. Real experimental conditions generally lie between these two idealized limits, depending on factors such as sample size and geometry, electrodes, and mounting constraints.}  The electric field was then removed and the system was further relaxed for $8.0\times10^{3}$ time steps to obtain the equilibrium domain configuration at zero field.

\medskip
\textbf{Supporting Information} \par 
Supporting Information is available from the Wiley Online Library or from the author.

\medskip
\textbf{Acknowledgements} \par 

The work was supported by the National Science Foundation under Grants No. DMR-2133373. L.Q.C. also appreciates the generous support from the Donald W. Hamer Foundation through a Hamer Professorship at Penn State.

\medskip
\textbf{Conflicts of Interest} \par 

L.Q.C. has a financial interest in MuPRO, LLC, a company which licenses and markets the software package used in this research.

\medskip

%


\begin{thebibliography}{10}
\providecommand{\url}[1]{\texttt{#1}}
\providecommand{\urlprefix}{URL }

\bibitem{park1997ultrahigh}
S.-E. Park, T.~R. Shrout,
\newblock \emph{Journal of Applied Physics} \textbf{1997}, \emph{82}, 4 1804.

\bibitem{kim2024electrical}
H.-P. Kim, M.-H. Zhang, B.~Wang, H.~Wu, Z.~Xu, S.~Liu, S.~Moon, Y.~Yamashita,
  J.~E. Ryu, J.~Liu, S.~Zhang, L.-Q. Chen, X.~Jiang,
\newblock \emph{Nature Communications} \textbf{2024}, \emph{15}, 1 6420.

\bibitem{sun2021dielectric}
Y.~Sun, T.~Karaki, Z.~Wang, T.~Fujii, Y.~Yamashita,
\newblock \emph{Japanese Journal of Applied Physics} \textbf{2021}, \emph{60},
  SF SFFC04.

\bibitem{chang2018dielectric}
W.-Y. Chang, C.-C. Chung, C.~Luo, T.~Kim, Y.~Yamashita, J.~L. Jones, X.~Jiang,
\newblock \emph{Materials Research Letters} \textbf{2018}, \emph{6}, 10 537.

\bibitem{qiu2020transparent}
C.~Qiu, B.~Wang, N.~Zhang, S.~Zhang, J.~Liu, D.~Walker, Y.~Wang, H.~Tian, T.~R.
  Shrout, Z.~Xu, et~al.,
\newblock \emph{Nature} \textbf{2020}, \emph{577}, 7790 350.

\bibitem{liu2024large}
X.~Liu, F.~Li, J.~Li, P.~Tan, J.~Lei, W.~Zhao, K.~Song, H.~Zheng, H.~Tian,
  F.~Li, X.~Wei, Z.~Xu,
\newblock \emph{Advanced Optical Materials} \textbf{2024}, \emph{12}, 33
  2400885.

\bibitem{cheng2017demonstration}
H.~Cheng, J.~Ouyang, Y.-X. Zhang, D.~Ascienzo, Y.~Li, Y.-Y. Zhao, Y.~Ren,
\newblock \emph{Nature Communications} \textbf{2017}, \emph{8}, 1 1999.

\bibitem{tyagi2026pressure}
S.~Tyagi, P.~C. Rout, S.~Singh, U.~Schwingenschl{\"o}gl,
\newblock \emph{Advanced Electronic Materials} \textbf{2026}, \emph{12}, 2
  e00230.

\bibitem{singh2025v2se2o}
S.~Singh, P.~C. Rout, M.~Ghadiyali, U.~Schwingenschl{\"o}gl,
\newblock \emph{Materials Science and Engineering: R: Reports} \textbf{2025},
  \emph{166} 101017.

\bibitem{zhang2012high}
S.~Zhang, F.~Li,
\newblock \emph{Journal of Applied Physics} \textbf{2012}, \emph{111}, 3
  031301.

\bibitem{liu2022ferroelectric}
X.~Liu, P.~Tan, X.~Ma, D.~Wang, X.~Jin, Y.~Liu, B.~Xu, L.~Qiao, C.~Qiu,
  B.~Wang, W.~Zhao, C.~Wei, K.~Song, H.~Guo, X.~Li, S.~Li, X.~Wei, L.-Q. Chen,
  Z.~Xu, F.~Li, H.~Tian, S.~Zhang,
\newblock \emph{Science} \textbf{2022}, \emph{376}, 6591 371.

\bibitem{finkel2022simultaneous}
P.~Finkel, M.~G. Cain, T.~Mion, M.~Staruch, J.~Kolacz, S.~Mantri, C.~Newkirk,
  K.~Kavetsky, J.~Thornton, J.~Xia, M.~Currie, T.~Hase, A.~Moser, P.~Thompson,
  C.~A. Lucas, A.~Fitch, J.~M. Cairney, S.~D. Moss, A.~G.~A. Nisbet, J.~E.
  Daniels, S.~E. Lofland,
\newblock \emph{Advanced Materials} \textbf{2022}, \emph{34}, 7 2106827.

\bibitem{negi2023ferroelectric}
A.~Negi, H.~P. Kim, Z.~Hua, A.~Timofeeva, X.~Zhang, Y.~Zhu, K.~Peters,
  D.~Kumah, X.~Jiang, J.~Liu,
\newblock \emph{Advanced Materials} \textbf{2023}, \emph{35}, 22 2211286.

\bibitem{xu2026cryogenic}
Z.~Xu, U.~Eckstein, Z.~Wang, H.~Wan, A.~Negi, A.~Naderi, J.~Liu, K.~G. Webber,
  X.~Jiang,
\newblock \emph{Journal of Applied Physics} \textbf{2026}, \emph{139}, 8.

\bibitem{otonicar2018multiscale}
M.~Otonicar, H.~Ursic, M.~Dragomir, A.~Bradesko, G.~Esteves, J.~Jones,
  A.~Bencan, B.~Malic, T.~Rojac,
\newblock \emph{Acta Materialia} \textbf{2018}, \emph{154} 14.

\bibitem{kim2022areview}
H.-P. Kim, H.~Wan, C.~Luo, Y.~Sun, Y.~Yamashita, T.~Karaki, H.-Y. Lee,
  X.~Jiang,
\newblock \emph{IEEE Transactions on Ultrasonics, Ferroelectrics, and Frequency
  Control} \textbf{2022}, \emph{69}, 11 3037.

\bibitem{maiwa2025microstructure}
H.~Maiwa, Y.~Xiang, Y.~Sun, H.-Y. Lee, Y.~J. Yamashita,
\newblock \emph{Microstructures} \textbf{2025}, \emph{5}, 3.

\bibitem{sun2022recent}
Y.~Sun, T.~Karaki, Y.~Yamashita,
\newblock \emph{Japanese Journal of Applied Physics} \textbf{2022}, \emph{61},
  SB SB0802.

\bibitem{qiu2021insitu}
C.~Qiu, Z.~Xu, Z.~An, J.~Liu, G.~Zhang, S.~Zhang, L.-Q. Chen, N.~Zhang, F.~Li,
\newblock \emph{Acta Materialia} \textbf{2021}, \emph{210} 116853.

\bibitem{liu2019realtime}
Y.~Liu, J.~Xia, P.~Finkel, S.~D. Moss, X.~Liao, J.~M. Cairney,
\newblock \emph{Acta Materialia} \textbf{2019}, \emph{175} 436.

\bibitem{sato2024response}
Y.~Sato,
\newblock \emph{Applied Physics Letters} \textbf{2024}, \emph{125}, 24 242906.

\bibitem{zhang2023varied}
D.~Zhang, L.~Wang, L.~Li, P.~Sharma, J.~Seidel,
\newblock \emph{Microstructures} \textbf{2023}, \emph{3}, 4 N/A.

\bibitem{perez-moyet2023anisotropy}
R.~P{\'e}rez-Moyet, Y.~Cardona-Quintero, I.~M. Doyle, A.~A. Heitmann,
\newblock \emph{Journal of the American Ceramic Society} \textbf{2023},
  \emph{106}, 9 5522.

\bibitem{sun2025enhanced}
J.-W. Sun, A.~Fluerasu, Z.~Xu, S.~Liu, S.-G. Lee, W.~Jo, X.~Jiang, J.~E. Ryu,
\newblock \emph{Journal of the American Ceramic Society} \textbf{2025}, e20610.

\bibitem{brichta2025quantifying}
N.~M. Brichta, L.~Tegg, L.~W. Giles, J.~E. Daniels, J.~M. Cairney,
\newblock \emph{Journal of the American Ceramic Society} \textbf{2025},
  \emph{108}, 5 e20346.

\bibitem{ushakov2021domain}
{\relax AD}.~Ushakov, Q.~Hu, X.~Liu, Z.~Xu, X.~Wei, V.~Y. Shur,
\newblock \emph{Applied Physics Letters} \textbf{2021}, \emph{118}, 23.

\bibitem{yang2024first}
F.~Yang, L.~Chen,
\newblock \emph{arXiv preprint arXiv:2412.04308} \textbf{2024}.

\bibitem{yang2025terahertz}
F.~Yang, X.~Li, D.~Talbayev, L.~Chen,
\newblock \emph{Physical review letters} \textbf{2025}, \emph{135}, 5 056901.

\bibitem{wan2021alternating}
H.~Wan, C.~Luo, C.~Liu, W.-Y. Chang, Y.~Yamashita, X.~Jiang,
\newblock \emph{Acta Materialia} \textbf{2021}, \emph{208} 116759.

\bibitem{xiong2022performance}
J.~Xiong, Z.~Wang, X.~Yang, R.~Su, W.~Zhang, X.~Long, Y.~Liu, C.~He,
\newblock \emph{Scripta Materialia} \textbf{2022}, \emph{215} 114694.

\bibitem{krogstad2018relation}
M.~J. Krogstad, P.~M. Gehring, S.~Rosenkranz, R.~Osborn, F.~Ye, Y.~Liu,
  J.~P.~C. Ruff, W.~Chen, J.~M. Wozniak, H.~Luo, O.~Chmaissem, Z.-G. Ye,
  D.~Phelan,
\newblock \emph{Nature Materials} \textbf{2018}, \emph{17} 718.

\bibitem{eremenko2019local}
M.~Eremenko, V.~Krayzman, A.~Bosak, H.~Y. Playford, K.~W. Chapman, J.~C.
  Woicik, B.~Ravel, I.~Levin,
\newblock \emph{Nature Communications} \textbf{2019}, \emph{10} 2728.

\bibitem{zheng2024heterogeneous}
H.~Zheng, T.~Zhou, D.~Sheyfer, J.~Kim, J.~Kim, T.~D. Frazer, Z.~Cai, M.~V.
  Holt, Z.~Zhang, J.~F. Mitchell, L.~W. Martin, Y.~Cao,
\newblock \emph{Science} \textbf{2024}, \emph{384}, 6703 1447.

\bibitem{bokov2026domain}
A.~A. Bokov, H.~Guo, M.~Bari, Z.-G. Ye,
\newblock \emph{Acta Materialia} \textbf{2026}, \emph{309} 122095.

\bibitem{perezmoyet2026insitu}
R.~P{\'e}rez-Moyet, W.~A. Kaminsky, Y.~Cardona-Quintero, W.~A. Visser, A.~A.
  Heitmann, H.~C. Robinson,
\newblock \emph{Journal of the American Ceramic Society} \textbf{2026},
  \emph{109}, 3 e70643.

\bibitem{luo2020multilayered}
C.~Luo, W.-Y. Chang, M.~Gao, C.-H. Chang, J.~Li, D.~Viehland, J.~Tian,
  X.~Jiang,
\newblock \emph{Acta Materialia} \textbf{2020}, \emph{182} 10.

\bibitem{patel2015mechanical}
S.~Patel, A.~Chauhan, R.~Vaish,
\newblock \emph{Journal of Applied Physics} \textbf{2015}, \emph{117}, 8
  084102.

\bibitem{bencan2020domain}
A.~Bencan, G.~Drazic, H.~Ursic, M.~Makarovic, M.~Komelj, T.~Rojac,
\newblock \emph{Nature communications} \textbf{2020}, \emph{11}, 1 1762.

\bibitem{schultheiss2023ferroelectric}
J.~Schulthei{\ss}, G.~Picht, J.~Wang, Y.~Genenko, L.~Chen, J.~Daniels,
  J.~Koruza,
\newblock \emph{Progress in Materials Science} \textbf{2023}, \emph{136}
  101101.

\bibitem{sun2014relaxorbased}
E.~Sun, W.~Cao,
\newblock \emph{Progress in Materials Science} \textbf{2014}, \emph{65} 124.

\bibitem{jablonski2016asymmetric}
M.~L. Jablonski, S.~Liu, C.~R. Winkler, A.~R. Damodaran, I.~Grinberg, L.~W.
  Martin, A.~M. Rappe, M.~L. Taheri,
\newblock \emph{ACS Applied Materials \& Interfaces} \textbf{2016}, \emph{8}, 5
  2935.

\bibitem{zhang2019intrinsic}
Y.~Zhang, H.~Lu, X.~Yan, X.~Cheng, L.~Xie, T.~Aoki, L.~Li, C.~Heikes, S.~P.
  Lau, D.~G. Schlom, L.~Chen, A.~Gruverman, X.~Pan,
\newblock \emph{Advanced Materials} \textbf{2019}, \emph{31}, 36 1902099.

\bibitem{mervosh2026far}
M.~W. Mervosh, C.~A. Randall,
\newblock \emph{Acta Materialia} \textbf{2026}, \emph{302} 121627.

\bibitem{mervosh2025far}
M.~W. Mervosh, J.-H. Lee, C.~A. Randall,
\newblock \emph{Nano Convergence} \textbf{2025}, \emph{12}, 1 62.

\bibitem{chen2008phase}
L.-Q. Chen,
\newblock \emph{Journal of the American Ceramic Society} \textbf{2008},
  \emph{91}, 6 1835.

\bibitem{wang2019understanding}
J.-J. Wang, B.~Wang, L.-Q. Chen,
\newblock \emph{Annual Review of Materials Research} \textbf{2019}, \emph{49},
  1 127.

\bibitem{khakpash2015misfit}
N.~Khakpash, H.~Khassaf, G.~Rossetti, S.~Alpay,
\newblock \emph{Applied Physics Letters} \textbf{2015}, \emph{106}, 8.

\bibitem{hlinka2007mobility}
J.~Hlinka,
\newblock \emph{Ferroelectrics} \textbf{2007}, \emph{349}, 1 49.

\end{thebibliography}
\end{document}


\title{Supporting Information}
\author{Yuan-Jie Sun}
\affiliation{Department of Materials Science and Engineering and Materials Research Institute, The Pennsylvania State University, University Park, PA 16802, USA}
\author{Bo Wang}
\email{bzw133@psu.edu}
\affiliation{Department of Materials Science and Engineering and Materials Research Institute, The Pennsylvania State University, University Park, PA 16802, USA}
\author{Long-Qing Chen}
\email{lqc3@psu.edu}
\affiliation{Department of Materials Science and Engineering and Materials Research Institute, The Pennsylvania State University, University Park, PA 16802, USA}
\maketitle

\FloatBarrier
\section{Energy comparison}
\begin{table}[!htbp]
\centering
\caption{Energy density differences (unit: J/(m$^3$)) between the final relaxed state and the initial relaxed state after poling for different $w/h$. The case $w/h=0.74$ corresponds to domain-wall elimination, whereas $w/h=0.92$ corresponds to the non-elimination case.}
\label{tab:energy_density_wh}

\small
\setlength{\tabcolsep}{6pt}
\renewcommand{\arraystretch}{1.15}

\begin{tabular}{c c c c c c c}
\hline
Initial $w/h$ & $E_{\text{initial}}$ &
$\Delta E_{\text{elastic}}$ & $\Delta E_{\text{electric}}$ &
$\Delta E_{\text{Landau}}$ & $\Delta E_{\text{gradient}}$ &
$\Delta E_{\text{total}}$ \\
\hline
0.74 & $-4.29\times10^{4}$ & $-347$ & $-3.62$ & $-307$ & $-403$ & $-1060$ \\
0.92 & $-4.29\times10^{4}$ & $-26.4$ & $-1.70$ & $\phantom{-}14.0$ & $\phantom{-}199$ & $\phantom{-}184$ \\
\hline
\end{tabular}
\end{table}

\renewcommand{\thefigure}{S\arabic{figure}}
\setcounter{figure}{0}

\begin{figure}[!htbp]
  \centering
  \includegraphics[width=0.8\textwidth]{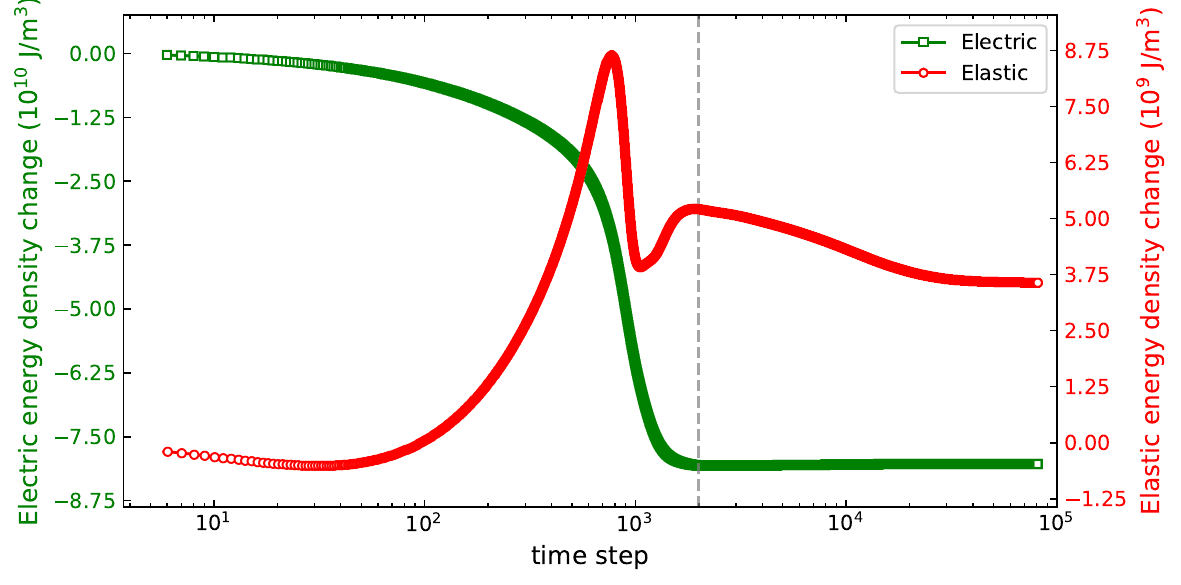}
  \caption{Temporal evolution of the electric and elastic energy contributions during poling of $w/h=0.92$ case.}
\end{figure}

\FloatBarrier
\section{Multi-cycle poling process}
\begin{figure}[!htbp]
  \centering
  \includegraphics[width=0.7\textwidth]{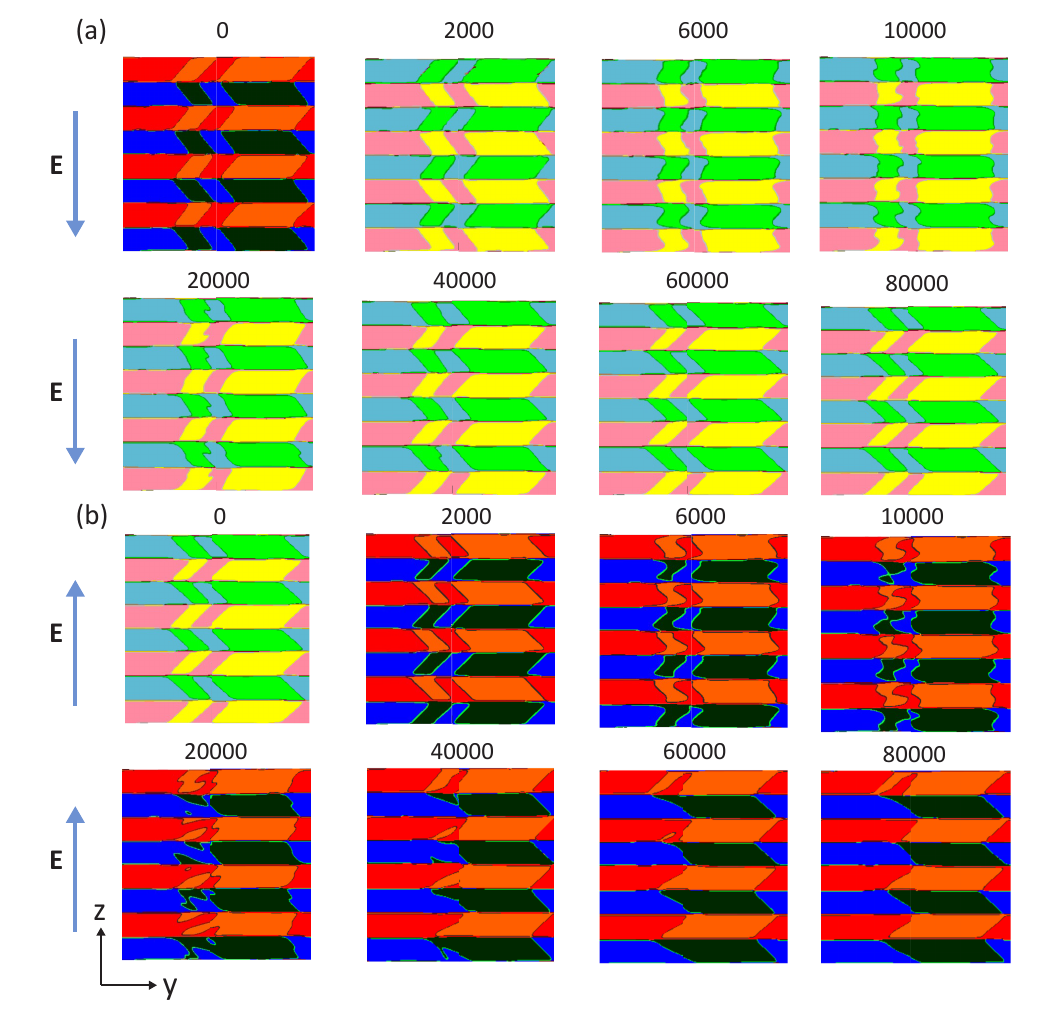}
  \caption{Multi-cycle poling process for the $w/h=0.94$ case. (a) Poling with an electric field along $[00\bar{1}]$ yields polarization switching without domain-wall elimination. (b) Continued poling along $[001]$ subsequently drives domain-wall elimination and produces the eliminated-wall configuration.}
\end{figure}

\newpage
\FloatBarrier
\section{Simulations of slab sample}
\begin{figure}[!htbp]
  \centering
  \includegraphics[width=0.7\textwidth]{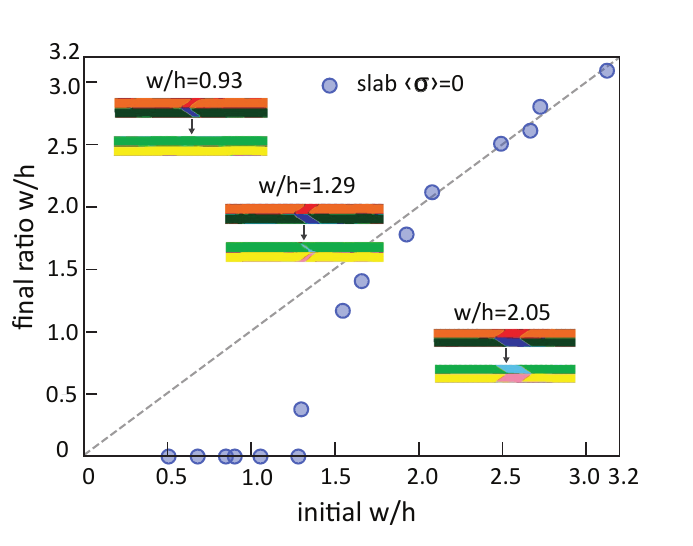}
  \caption{Domain size analysis of a single slab similar to Fig. 4 with mechanical boundary condition $\langle\boldsymbol{\sigma}\rangle=0$. All other boundary conditions are identical. The domain height is $h=32$ nm and the total length of the slab is 512 nm.}
\end{figure}



\FloatBarrier
\section{Asymmetry factor}
\begin{figure}[!htbp]
  \centering
  \includegraphics[width=0.4\textwidth]{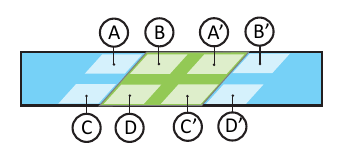}
  \caption{Schematic drawing of the sampling regions (A--D and A$'$--D$'$) used to evaluate the asymmetry factor.}
  \label{fig:asymmetry}
\end{figure}
As shown in main text, the poling-induced domain dynamics is dominated by elastic interactions: the elastic driving force controls both the domain-wall migration and the subsequent elimination process. The elastic driving force is defined as:
\begin{equation}
D_{i,\text{elast}}=-\frac{\partial E_{\text{elast}}}{\partial P_i},
\label{eq:SI1}
\end{equation}
where $i$ denotes the Cartesian direction. In the present case, the polarization discontinuity across the 71$^\circ$ domain wall is along the $x$ direction, so the relevant term reduces to $D_{x,\text{elast}}$.

We use four regions with concentrated elastic driving force adjacent to a 71$^\circ$ domain wall labeled $A$ to $D$ in the inset of Fig.~\ref{fig:asymmetry} to track the dynamics of an individual wall. An imbalance in the elastic driving force between the upper and lower wall segments drives a swinging reorientation of the wall, manifested as a gradual rotation of its local orientation. We therefore define an imbalanced driving-force metric as a quantitative proxy for the swing velocity of the domain wall:

\begin{equation}
    I_{\mathrm{wall}}= \big| (|\langle D_{x,\mathrm{elas}}^{B} \rangle| - |\langle D_{x,\mathrm{elas}}^{A} \rangle | )
                   - ( |\langle D_{x,\mathrm{elas}}^{D} \rangle| - |\langle D_{x,\mathrm{elas}}^{C} \rangle| )\big|, \label{eq:wall}
\end{equation}
where $\langle D_{x,\mathrm{elas}}^{X}\rangle$ ($X=A,B,C,D$) denotes the spatially averaged elastic driving force within region $X$, as defined in Fig.~\ref{fig:asymmetry}. All analysis windows are selected from either the upper or lower one-third of the layer thickness $h$ and use an identical lateral width of $h/2$. This metric quantifies the extent to which the elastic driving force biases reorientation of the 71$^\circ$ domain wall orientations during AC poling. A larger magnitude of the imbalanced driving force corresponds to a higher domain-wall swing velocity.

Although swinging reorientation is observed for essentially all 71$^\circ$ domain walls, it does not necessarily culminate in wall elimination. Whether elimination occurs is primarily determined by the degree of asymmetric motion between adjacent walls. To quantify the propensity for domain elimination, we introduce an asymmetry factor for a pair of neighboring domains, denoted as $\mathcal{A}_{\mathrm{domain}}$, defined as:
\begin{equation}
    \mathcal{A}_{\mathrm{domain}}= \big| I_{\mathrm{wall}} - I_{\mathrm{wall}}' \big|, \label{eq:domain}
\end{equation}
where $I_{\mathrm{wall}}'$ is defined similarly as $I_{\mathrm{wall}}$, but evaluated using the corresponding regions $X'$. A larger value of $\mathcal{A}_{\mathrm{domain}}$ indicates a greater mismatch in the swing velocities of the two adjacent domain walls. Such an enhanced asymmetry results in correlated domain wall contact and the formation of circular domain islands. These islands subsequently shrink and vanish, leading to domain wall elimination.

\begin{figure}[!htbp]
  \centering
  \includegraphics[width=0.55\textwidth]{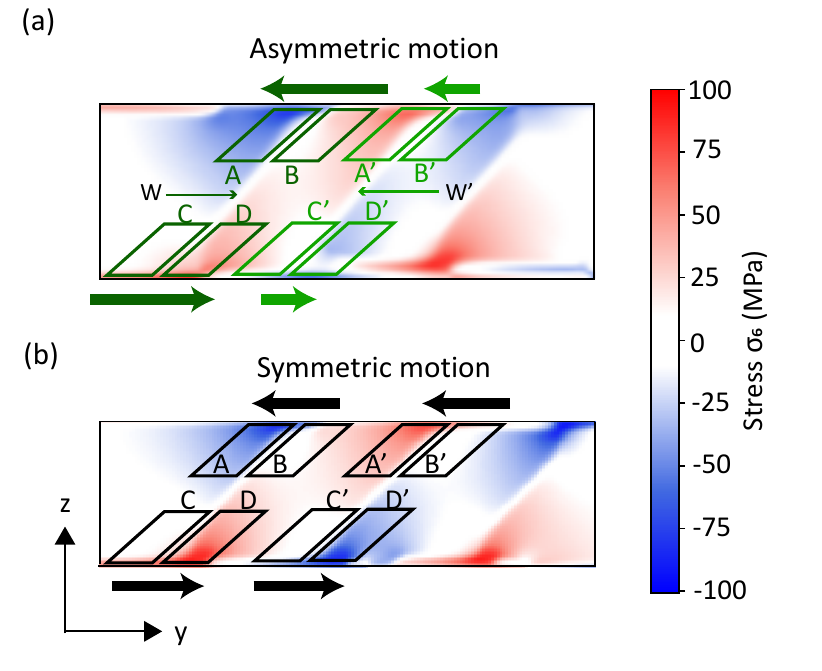}
  \caption{Local shear-stress fields $\sigma_6$ ($\sigma_{xy}$) around adjacent 71$^\circ$ domain walls. (a) Correlated walls exhibit different local $\sigma_6$ environments in regions $A$--$D$ and $A'$--$D'$. (b) Uncorrelated walls exhibit more similar local $\sigma_6$ environments and undergo nearly symmetric motion.}
  \label{fig:stress_local}
\end{figure}

 {To better illustrate the origin of the asymmetric elastic driving force, we plot the local shear-stress component $\sigma_6$ ($\sigma_{xy}$) for correlated domain-wall pairs in Figure~S5(a) and uncorrelated domain-wall pairs in Figure~S5(b). In the correlated case, the two adjacent 71$^\circ$ walls, denoted as $W$ and $W'$, experience different local elastic environments. Specifically, wall $W$ exhibits a stronger stress contrast across the wall than wall $W'$, indicating a larger local elastic driving force. In contrast, in the uncorrelated case shown in Figure~S5(b), the two walls experience similar local stress environments and therefore undergo nearly symmetric motion. We note that the shear-stress map is intended only as a visual illustration of the local elastic-field imbalance. A quantitative evaluation of the elastic driving-force asymmetry requires calculating $D_{x,\mathrm{elast}}=-\partial E_{\mathrm{elast}}/\partial P_x$ using the sampling regions defined in Figure~S4, as described in Equations~1--3. The resulting asymmetry factor is shown in Figure~4(b).}

\section{Electrostatic asymmetry analysis}

 {
To isolate the energetic origin of the symmetry breaking, we performed a control analysis in which the asymmetry factor was evaluated using the local electric driving force rather than the elastic driving force:
\begin{equation}
    D_{i,\mathrm{elec}}=-\frac{\partial E_{\mathrm{elec}}}{\partial P_i}
\end{equation}
The analysis was carried out using the identical direction along $x$ ($D_{x,\mathrm{elec}}$), sampling regions, time window, and definition adopted for the elastic asymmetry factor (SI Section IV).
}

\begin{figure}[h!]
    \centering
    \includegraphics[width=0.6\linewidth]{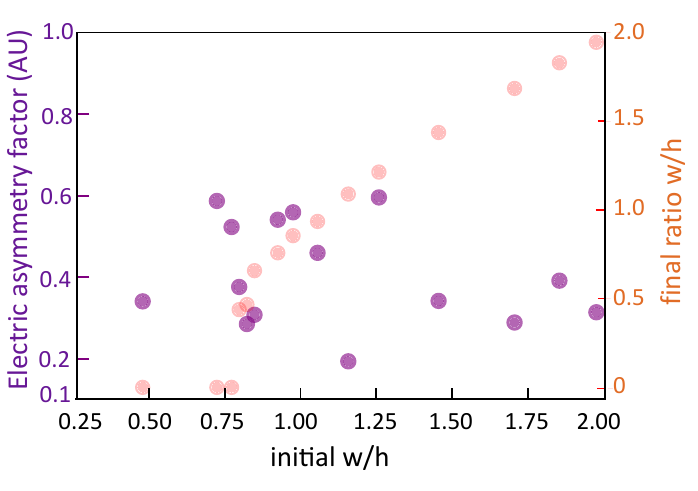}
    \caption{ Dependence of the electric asymmetry factor (left axis) on $w/h$, plotted together with the resulting $w/h$ under clamping mechanical boundary condition. }
    \label{fig:electric_asym}
\end{figure}

 {
As shown in the Figure~\ref{fig:electric_asym}, in contrast to the elastic asymmetry factor, the electric asymmetry factor exhibits no systematic dependence on $w/h$ and no correlation with domain-wall elimination. This comparison indicates that the geometry-dependent imbalance responsible for the asymmetric motion of neighboring 71$^\circ$ domain walls is dominated by the elastic driving force rather than the electrostatic contribution.
}
